\documentclass[a4paper,12pt]{article}
\usepackage{tabularx}
\usepackage{longtable, float}
\include{jm-tex-macros-public}
\usepackage{cite}
\usepackage[utf8]{inputenc}
\usepackage{subfigure}
\usepackage{mathtools}
\usepackage{amssymb}
\usepackage{amsmath}

\renewcommand{\title}[1]{\vbox{\center\LARGE{#1}}\vspace{5mm}}
\renewcommand{\author}[1]{\vbox{\center#1}\vspace{5mm}}
\newcommand{\address}[1]{\vbox{\center\em#1}}
\newcommand{\email}[1]{\vbox{\center\tt#1}\vspace{5mm}}

\renewcommand{\date}[1]{\vbox{\center#1}}

\begin{document}

\title{Finite temperature corrections to black hole quasinormal modes from 2D CFT} 
\author{Sanchari Pal} 
\email{pals@iitk.ac.in}

\address{    {\it 
Department of Physics, 
 Indian Institute of Technology - Kanpur, \\
Kanpur 208016, India\\
 }}

\abstract{We revisit the holographic calculation of the decay modes of the BTZ blackhole perturbed by a scalar probe. We carry out a finite temperature expansion of the torus two point function of large central charge $(c)$ CFTs in order to obtain the corrections to blackhole quasinormal modes. We take the contribution of the lightest primary above the vacuum, with dimension $\Delta_{\chi}$ and upper bound of $c/6$, in the torus two point function. We exploit the modular properties of 2D CFT on the torus to connect the expression of thermal two point function at high temperature with the same at low temperature. The correction term can be expressed as a four point function on the Riemann sphere. In the dual bulk theory the corrections are expected to arise due to the presence of a light matter field in the spacetime along with gravity. In the holographic limit the effects of this correction term is encoded in the change in blackhole temperature. This leads to new quasinormal modes and modification of thermalization time scale.}

\tableofcontents

\vfill\eject

\section{\label{sec:intro}Introduction}Whenever a blackhole is perturbed with a probe\footnote{The  probe field is not heavy enough to change the black hole global temperature.} the perturbation propagates with certain discrete frequencies due to fixed boundary conditions at horizon and asymptotic infinity\cite{Birmingham:2001hc}. The flux has to be ingoing at the blackhole horizon and outgoing at the asymptotic infinity. The frequencies depend upon the parameters of the probe and the black hole.  For asymptotically flat space-time there have been many works in the gravitational context starting from the 1970s\cite{Vishveshwara:1970zz,Chandrasekhar:1975zza, Kokkotas:1999bd}. The calculation of the quasinormal modes (QNM) follows from solving the field wave equations of the probe in the black hole background with suitable boundary conditions, and then calculating the emission/absorption cross-section using classical scattering theory. \vspace{2mm}\\ 
Brown and Henneaux \cite{Brown:1986nw} showed that in three dimensions the asymptotic symmetries of Einstein gravity with negative cosmological constant are  described by two copies of Virasoro algebra. It was realised in \cite{Maldacena:1997re}, \cite{Witten:1998qj} and \cite{Gubser:1998bc} that quantum gravity in AdS $d+1$ dimensions is dual to Conformal Field Theory (CFT) in d dimensions, living on the boundary of AdS. Even before this duality was proposed it was found analytically \cite{Maldacena:1997ih} that in both four and five dimensions the semiclassical emission rates of a blackhole perturbed by a scalar with orbital angular momentum agree, in striking detail, with the effective conformal field theory calculations in one less dimensions, without the theory having to take recourse of fundamental string theory calculations.   
\begin{align}
    \Gamma^{(decay)}\propto e^{-\frac{\omega}{2T_{H}}}\omega^{2l-1} T_{H}^{2l+1} \bigg\vert\Gamma\bigg(1+l+\frac{i\omega}{2\pi T_{H}}\bigg)\bigg\vert^{2}
\end{align}
where on the right hand side $\omega$ is the frequency of decay of perturbation, $l$ is the angular momentum of the probe scalar and $T_{H}$ is the blackhole temperature. Carrying forward this conjecture, an interpretation of the blackhole quasinormal modes in terms of dual thermal CFT was first suggested in \cite{Horowitz:1999jd}. The analysis of \cite{Horowitz:1999jd} involves numerically calculating quasinormal modes for Schwarzschild blackhole in asymptotically AdS space-time.  \\
Analytic computations for decay rate of blackhole perturbations have been carried out in the BTZ background in \cite{Birmingham:1997rj}. The BTZ blackhole metric is a solution of vacuum Einstein field equations in three dimensions with negative cosmological constant, $\lambda=-1/L^{2}$ \cite{PhysRevLett.69.1849} where $L$ is the radius of curvature of the space-time. The exact computation of the BTZ blackhole probed by a mass-less scalar field yields the following decay rate
\begin{equation}
\label{decay1}
\Gamma^{(\text{decay})} \propto e^{-\frac{\omega}{2T_{H}}} \omega^{-1} T_{H}  \bigg\vert \Gamma\left( 1  +  i \frac{\omega}{4\pi T_{L}} \right) \Gamma\left(1+i\frac{\omega}{4\pi T_{R}}\right)\bigg\vert^2,
\end{equation}
where $T_{L/R}^{-1}=T_{H}^{-1}\left(1\pm\frac{r_{-}}{r_{+}}\right)$ and $r_{\mp}$ are the inner and outer horizons of the blackhole. \\ 
Probing the blackhole is equivalent to perturbing this thermal state and the imaginary part of the frequency of decay determines the thermalization time-scale. From bulk perspective a prediction of the thermalization time-scale was first obtained in \cite{Horowitz:1999jd}. On the boundary the scalar probe is coupled to a operator $\mathcal{O}$ with dimension $\Delta_{\mathcal{O}}$ in 2D CFT. Completely bypassing the string theoretic approach as espoused in \cite{Birmingham:1997rj}, it can be shown  \cite{Birmingham:2001pj} that the location of the poles of the fourier transformation of the high temperature conformal autocorrelator of $\mathcal{O}$ exactly matches with the frequencies of QNMs obtained from bulk side calculations. 
\begin{equation}
\label{decay2}
    \Gamma^{(decay)}\propto \frac{e^{-\omega\beta/2}}{\beta^{2\Delta_{\mathcal{O}}-1}} \bigg\vert\Gamma\left(\Delta_{\mathcal{O}}+i\frac{\omega\beta}{2\pi}\right)\bigg\vert^{2}  
\end{equation}
\begin{equation} 
   \label{qnm_mode}
    \omega_{n}=-i\frac{2\pi}{\beta}(n+\Delta_{\mathcal{O}})
 \end{equation} 
 Where $\beta$ is the inverse temperature of the blackhole. 
The mass-less scalar probe in \cite{Birmingham:1997rj} has mass dimension $\Delta_{\mathcal{O}}=1$.\\ 
The thermal two point function on torus does not have a closed form. At low temperature limit $\beta\to\infty$ it can be approximated by two point function on a spatial cylinder. For our study we shall be interested in keeping the temperature finite and non-zero. We exploit the modular properties of the torus and relate the two point function at high temperature with the low temperature expression. To arrive at the decay rate of quasinormal modes \eqref{decay2} one only needs to deal with the vacuum contribution to the torus two point function. The rest of the terms are neglected as they are are exponentially $(e^{-4\pi^2/\beta})$ suppressed. The expression for quasi-normal modes one obtains from the singularities of gamma function in eq. \eqref{decay2} is for pure gravity.  In this paper we take contributions from the low lying excited state $\vert\chi\rangle$ which adds correction to the high temperature approximation.\\             
 Recent bounds from modular bootstrap \cite{Hellerman:2009bu,Friedan:2013cba, Qualls:2013eha} indicate that at large $c$, the lightest non-vacuum primary has an upper bound $\frac{c+\bar{c}}{12}$. Using the AdS/CFT dictionary this bound translates to $1/4G_{N}$ which is the mass of the lightest blackhole. This indicates that incorporating corrections from low lying $\vert\chi\rangle$ states might be same as considering light matter in the bulk along with gravity.\\
In particular, in this paper we seek answers to the following questions
\begin{itemize}
\item  \textit{How does the decay rate as obtained in eq\eqref{decay2} change as we consider finite temperature corrections to the torus two point function?}
\item \textit{ Does the change lead to new quasi-normal modes?}
\end{itemize} 
We shall show that at large $c$ limit the correction to the decay rate are manifested as correction to the blackhole temperature. Eventually this will lead to shifts in quasi-normal modes.

\subsection*{Outline }
Section \ref{prelim} is a review of the mathematical tool that is used to calculate the correction to the quasinormal modes. We arrive at the amplitude of perturbation process (eq. \eqref{limit}) by coupling the scalar probe to CFT operator $\mathcal{O}$. We briefly discuss how the modular property of torus is exploited to relate the high temperature two point function with the low temperature one. Finally we arrive at the finite temperature correction to torus two point function. Having established the correction term, In section \ref{calculation} we evaluate it by conformally transforming the answer derived in \cite{Fitzpatrick:2014vua} through monodromy method. In section \ref{corrc} we discuss what this correction means in terms of changes in QNMs. In the concluding section (sec. \ref{discussion})  we discuss possible explanations of our results.\\
In \S \ref{ap:leading-high-temp} we derived the two point function on spatial cylinder. We also tried to figure out the non-identity block contribution to the correction term in \S\ref{other-block}. Finally in \S\ref{ap:vacuum desc} we argue that the descendants of vacuum will not have any effect on the QNMs.

\section{Preliminaries}
\label{prelim}
\subsection{Quantum corrections to  quasinormal modes from CFT}
In this section we review the basic ingredients that go on to the calculation of the quasinormal modes from CFT$_2$. One is interested in the probability of the emission of a particle with energy $\omega$, from the process of perturbing the black hole by a scalar. The black hole state in equilibrium is described by a thermal state of the CFT (denoted by $\vert i\rangle$). The operator dual to the scalar in the CFT will be denoted by $\mathcal{O}$. Let us further call the CFT state $\vert f\rangle$ to which the thermal state ends up in, upon the perturbation. Therefore the amplitude of the process is, 
\begin{equation}
    \mathcal{M}_{\omega} \sim \int dt e^{-i \omega t } \langle f\vert \mathcal{O}(t) \vert{i}\rangle. 
\end{equation}
Squaring, summing over the final state $\vert f\rangle$ and using completeness and translational invariance of the CFT correlator we get, 
\begin{eqnarray}
\sum_f |\mathcal{M}|^2 _{\omega}&\sim&  \int_{-\infty}^{\infty} dt e^{-i \omega t } \langle i\vert \mathcal{O}^\dagger(t) \mathcal{O}(0)\vert i\rangle \nonumber \\
&=& \int_{-\infty}^{\infty} dt e^{-i \omega t } \frac{1}{Z(\beta)}   \langle\mathcal{O}^\dagger(t) \mathcal{O}(0)\rangle_\beta \label{formula}
\end{eqnarray}
At high temperature ($\beta\to 0$) even though the conformal field theory lives on a circle with circumference $L$, since $\beta/L\to 0$, it can be defined on an infinite line. The thermal state is then given by a cylinder with the Euclidean time direction compactified with periodicity $\beta$. The cylinder geometry, being conformally related to the plane, allows one to obtain the thermal correlator from a conformal transformation of the plane correlator answer. This leads to the following expression for eq. \eqref{formula} : 
\begin{equation} 
\label{limit}
\sum_f |\mathcal{M} |^2_{\omega} \sim \int_{-\infty}^{\infty} dt\, e^{-i \omega t }\left(  \frac{ \pi / \beta}{\sinh \frac{\pi t}{\beta} } \right)^{2\Delta_{\mathcal{O}}}. 
\end{equation}
Eq. \eqref{decay2} follows from eq. \eqref{limit} by a contour integral with appropriate $i \epsilon$ prescription \cite{Maldacena:1997ih}. For techniques involved in solving the integration in eq. \eqref{limit} see \cite{Brehm:2018ipf}.  

\subsection{Modular properties of torus}
A torus can be built by identifying the points on a complex plane. The fundamental building block of torus on a complex plane is defined by two complex numbers $\alpha_{1}$ and $\alpha_{2}$. Any given point $w$ is identified with another point $w+m\alpha_{1}+n\alpha_{2}$ where $m$ and $n$ are integers. Shape of the torus depends on the modular parameter $\tau=\frac{\alpha_{1}}{\alpha_{2}}$. However the same torus can be described by two different complex numbers $\beta_{1}$ and $\beta_{2}$ which are related to $\alpha_{1}$ and $\alpha_{2}$ in the following manner. 
\begin{equation}
    \begin{pmatrix}
    \beta_{1}\\
    \beta_{2} \end{pmatrix} = \begin{pmatrix}
    a & b\\
    c & d
\end{pmatrix} \begin{pmatrix}
\alpha_{1}\\
\alpha_{2}
\end{pmatrix}
\end{equation}
where $ad-bc\neq 0$ and $a,b,c,d\in Z $\\  
In terms of modular parameter this means that both $\tau$ and $\frac{a\tau+b}{c\tau+d}$ define the same torus. We shall be interested in a specific transformation that takes $\tau$ to $-\frac{1}{\tau}$. This is named as S transformation.           
\subsection{High temperature expansion}
The thermal states of 2D CFT are described on a torus of periodicity of $\beta$ along Euclidean time direction and of length $L$ along the space direction. When temperature is high ($\beta/L\to 0$) one can reduce the torus to a thermal cylinder since the length along the space direction is infinite as compared to time direction. For low temperature ($\beta/L\to\infty$) the theory lives on a spatial cylinder. We exploit the modular properties of torus to connect the correlators at two extreme ends of the temperature range. The partition function remains invariant under S transformation. 
\begin{equation}
\label{partition1}
    Z(\tau,\bar{\tau})=\text{Tr}[e^{2\pi i\tau(L_{0}-c/24)}e^{-2\pi i\bar{\tau}(\bar{L}_{0}-\bar{c}/24)}]
\end{equation}
Here $\tau=\frac{i\beta}{L}$. The trace is taken over all the primaries and their descendants. To simplify factors and to make our corrections sharper, we set $L = 2\pi$. Using modular invariance of partition function we get the high temperature partition function in terms of the low temperature one.
\begin{eqnarray}
Z(\tau)&=&Z(-1/\tau)\nonumber\\
  Z(\beta)&=&Z\left(\frac{4\pi^{2}}{\beta}\right)\nonumber\\
&=&e^{\frac{\pi^{2}c}{3\beta}}\text{Tr}[e^{-\frac{4\pi^{2}}{\beta}(L_{0}+\bar{L}_{0})}]\nonumber\\
&=&e^{\frac{\pi^{2}c}{3\beta}}\langle 0\vert e^{-\frac{4\pi^{2}}{\beta}(L_{0}+\bar{L}_{0})}\vert 0\rangle+e^{\frac{\pi^{2}c}{3\beta}}\sum_{N}g_{N}\langle N\vert e^{-\frac{4\pi^{2}}{\beta}(L_{0}+\bar{L}_{0})}\vert N\rangle\nonumber\\ 
&+&e^{\frac{\pi^{2}c}{3\beta}}\sum_{\chi}g_{\chi}\langle\chi\vert e^{-\frac{4\pi^{2}}{\beta}(L_{0}+\bar{L}_{0})}\vert\chi\rangle\nonumber\\  
&=&e^{\frac{\pi^{2}c}{3\beta^{2}}} \left(1+\sum_{N} g_{N}\hspace{1mm}e^{-\frac{4\pi^{2}}{\beta}N}  +\sum_{\chi} g_{\chi}\hspace{.5mm}e^{-\frac{4\pi^{2}}{\beta}\Delta_{\chi}}\right)\nonumber  
  \end{eqnarray}  
For $\beta\to 0$, $e^{-\frac{4\pi^{2}\Delta}{\beta}}$ is exponentially suppressed for higher values of $\Delta$. For consistency we need $\Delta_\chi \gtrsim \frac{c}{12}$. If $\Delta_\chi$ is too small then the excited states contribution would be significant. The state $\vert N\rangle$ denotes the descendants of vacuum of order $N$. It is given by $L_{-k_{1}}^{n_{1}}L_{-k_{2}}^{n_{2}}...L_{-k_{m}}^{n_{m}}\vert 0\rangle$ where $1\le k_{1}\le k_{2}...\le k_{m}$ and $\sum_{i} n_{i}k_{i}=N$. $\vert\chi\rangle$ is the state with degeneracy $g_\chi$ and includes non-vacuum primaries and their descendants. For rest of the paper we take $c = \bar{c}$. \\ 
The correlators transform with specific weights. Under s transformation the torus coordinate $w$ changes to $\frac{w}{\tau}$. A primary of weight $(h,\bar{h})$, transforms as 
\begin{equation}
\mathcal{O}_{h,\bar{h}}(w,\bar{w}) = \tau^{-h }\bar{\tau}^{-\bar{h}} \mathcal{O}_{h,\bar{h} }\left(\frac{w}{\tau}, \frac{\bar{w}}{\bar{\tau}}\right). \label{op-trafo}
\end{equation} 
Torus two point function undergoes the following transformation.
\begin{equation}
    \langle\mathcal{O}\left(w_{1}/\tau,\bar{w}_{1}/\bar{\tau}\right)\mathcal{O}\left(w_{2}/\tau,\bar{w}_{2}/\bar{\tau}\right)\rangle=\tau^{2h_{\mathcal{O}}}\bar{\tau}^{2\bar{h}_{\mathcal{O}}}\langle\mathcal{O}(w_{1},\bar{w}_{1})\mathcal{O}(w_{2},\bar{w}_{2})\rangle
\end{equation}
These properties of partition function and correlators can be successfully used to bootstrap the CFT data \cite{Nahm:1974jm,Cardy:1986ie,Kraus:2016nwo,Carlip:2000nv}.

%\textcolor{red}{Note : Look more closely to the sub-dominant terms. We can drop the second term since the descendants of identity would be dominant if we consider $\Delta_\chi \gtrsim \frac{c}{12}$ (?). In that case \eqref{towork} simplifies and will have `nicer' holographic interpretation. }
The low temperature ($\beta\to\infty$) expansion of the normalized two-point function at inverse temperature $\beta$ is,  
\begin{eqnarray}
\label{lowt}
\langle{\mathcal{O}(w_{1})\mathcal{O}(w_{2})}\rangle_{\beta}&=&\frac{1}{Z(\beta)} \text{Tr} \left(\mathcal{O}(w_{1})\mathcal{O}(w_{2})e^{-\beta(L_{0}-c/24)}e^{-\beta(\bar{L}_{0}-c/24)}\right)\nonumber \\
&=&\frac{1}{Z(\beta)}\bigg( \langle 0\vert\mathcal{O}(w_{1})\mathcal{O}(w_{2})\vert 0\rangle\exp(\beta c/12)\nonumber \\
&+&\sum_{N}g_{N}\langle N\vert\mathcal{O}(w_{1})\mathcal{O}(w_{2})\vert N\rangle e^{-\beta(N-c/12)}\nonumber\\
&+&\sum_\chi g_\chi \langle\chi\vert\mathcal{O}(w_{1})\mathcal{O}(w_{2})\vert\chi\rangle\exp^{-\beta(\Delta_{\chi}-c/12)}\bigg) 
\end{eqnarray}\\
The second term on the r.h.s of eq. \eqref{lowt} is due to the contribution from descendants of vacuum. $w$ is elliptic coordinate,
\begin{eqnarray}
w&=&t_{euc}+i\sigma\\
\bar{w}&=&t_{euc}-i\sigma
\end{eqnarray}
If it is a low temperature expansion ($\beta\to\infty$) we can neglect the terms with higher values of $\Delta_{\chi}$.  
The S modular transformation connects the low temperature expansion with the high temperature one. Since we are working in high temperature ($\beta \to 0$) limit, the correlator on the right hand side are to be calculated at  $\beta^{\prime}=\frac{4\pi^{2}}{\beta}\to\infty$ limit. This essentially means that the correlators are evaluated on spatial cylinder. 
\begin{eqnarray}
\langle\mathcal{O}(w_{1},\bar{w}_{1})\mathcal{O}(w_{2},\bar{w}_{2})\rangle_{\beta}&=&\tau^{-2h_{\mathcal{O}}}\bar{\tau}^{-2\bar{h}_{\mathcal{O}}}\langle\mathcal{O}(w_{1}^{\prime},\bar{w}_{1}^{\prime})\mathcal{O}(w_{2}^{\prime},\bar{w}_{2}^{\prime})_{(4\pi^{2}/\beta)} \\
\langle\mathcal{O}(w_{1},\bar{w}_{1})\mathcal{O}(w_{2},\bar{w}_{2})\rangle_{\beta}
&\approx& \tau^{-2h_{\mathcal{O}}}\bar{\tau}^{-2\bar{h}_{\mathcal{O}}}\langle 0\vert \mathcal{O}(w_{1}^{\prime},\bar{w}^{\prime}_{1})\mathcal{O}(w^{\prime}_{2},\bar{w}^{\prime}_{2})\vert 0\rangle e^{\frac{c \pi^{2}}{3 \beta}} \nonumber \\
&+&\tau^{-2h_{\mathcal{O}}}\bar{\tau}^{-2\bar{h}_{\mathcal{O}}}\sum_{N}g_{N}\hspace{1mm}\langle N\vert\mathcal{O}(w_{1}^{\prime},\bar{w}_{1}^{\prime})\mathcal{O}(w_{2}^{\prime},\bar{w}_{2}^{\prime})\vert N\rangle e^{-\frac{4\pi^{2}}{\beta}(N-\frac{c}{12})}\nonumber\\
&+&\sum_\chi g_\chi \tau^{-2h_{\mathcal{O}}}\bar{\tau}^{-2\bar{h}_{\mathcal{O}}} \langle\chi\vert \mathcal{O}(w_{1}^{\prime},\bar{w}^{\prime}_{1})\mathcal{O}(w^{\prime}_{2},\bar{w}^{\prime}_{2})\vert\chi\rangle e^{-\frac{4\pi^{2}}{\beta}(\Delta_{\chi}-\frac{c}{12})}. \nonumber\\ 
\end{eqnarray} 
We explain in appendix how the expression for contribution from the descendants of vacuum is reached at. \\  
We shall always choose $w_{2}$ be at $0$ ($\bar{w}_{2}=0$). Therefore, $w_{2}^{\prime}= \bar{w}_{2}^{\prime}= 0$. The other point $w_{1} = t_{euc}$ with the corresponding $\sigma_2$ being $0$. So, $\bar{w}_{1} = t_{euc}$. Later we shall be analytically continuing $t_{euc} = i t$ before performing the Fourier transform. 
We rewrite the thermal autocorrelator, 
\begin{eqnarray}
\langle\mathcal{O}(t_{euc})\mathcal{O}(0)\rangle_{\beta}&\approx&  
\left( \frac{\beta}{2\pi } \right)^{-2 \Delta_{\mathcal{O}} } i^{-2s_{\mathcal{O}}} \nonumber \\
 &&\bigg(
\frac{\langle 0\vert \mathcal{O}(t_{euc}/\tau)\mathcal{O}(0)\vert 0\rangle}{1+\sum_{N^{\prime}}g_{N^{\prime}}\hspace{.5mm} e^{-\frac{4\pi^{2}N^{\prime}}{\beta}}+\sum_{\chi'} g_{\chi'} e^{-\frac{4\pi^2 \Delta_{\chi'}}{\beta }}}  \nonumber \\
&+&\frac{\langle N\vert \mathcal{O}(t_{euc}/\tau)\mathcal{O}(0)\vert N\rangle \sum_{N}g_{N}\hspace{.5mm}e^{-\frac{4\pi^{2}N}{\beta}}}{1+\sum_{N^{\prime}}g_{N^{\prime}}\hspace{.5mm}  e^{-\frac{4\pi^{2}N^{\prime}}{\beta}}+\sum_{\chi'} g_{\chi'} e^{-\frac{4\pi^2 \Delta_{\chi'}}{\beta } }}  \nonumber \\
&+&\frac{\sum_{\chi} g_{\chi} \langle\chi\vert \mathcal{O}(t_{euc}/\tau)\mathcal{O}(0)\vert\chi\rangle e^{-\frac{4\pi^2 \Delta_{\chi}}{\beta } }}{1+\sum_{N^{\prime}}g_{N^{\prime}}\hspace{.5mm}  e^{-\frac{4\pi^{2}N^{\prime}}{\beta}}+\sum_{\chi'} g_{\chi'} e^{-\frac{4\pi^2 \Delta_{\chi'}}{\beta } }} \bigg), \label{towork}
\end{eqnarray}
where, spin $s_{\mathcal{O}} = h_{\mathcal{O}} - \bar{h}_{\mathcal{O}}$. \\
It is important to note that the leading term on r.h.s of \eqref{towork} leads to \eqref{limit}. The evaluation and reproduction of the standard leading result from $\langle\mathcal{O}(w_{1},\bar{w}_{1})\mathcal{O}(0,0)\rangle$ is put in \S\ref{ap:leading-high-temp}.The contribution from the descendants of vacuum does not affect the thermalization timescale or the frequencies of QNMs. In \S\ref{ap:vacuum desc} we prove this inductively. For the rest of the paper we shall concern ourselves with the evaluation of the last term.                

\section{Evaluation of Correction Term} 
\label{calculation}
%According to the $AdS/CFT$ correspondence the blackholes are described by micro-canonical ensemble of heavy primary states in dual CFT which appear thermal at large c limit \cite{Fitzpatrick:2014vua}. The blackhole temperature is related to the scaling dimension of the conformal operator via, 

\subsection{$c\to\infty$ and $h_{\mathcal{O}}/c\gg h_{\chi}/c$} 
In the last section, we have succeeded in writing the thermal correlator in terms of correlators on the spatial cylinder. To evaluate \eqref{towork} we need to compute a correlator on the spatial cylinder:
$\langle\chi\vert \mathcal{O}(0,0)\mathcal{O}(t_{euc}/\tau, t_{euc}/\bar{\tau})\vert\chi\rangle $. 
We shall evaluate it by going to the plane. For simplicity we only work with the holomorphic part, supplementing with the anti-holomorphic part only at the end. Also we assume that we only deal with scalar primaries (ignoring spin for now).
In the limit of large $c$ and the ratios $h_{\mathcal{O}}/c, h_{\chi}/c$ held fixed, with $h_{\mathcal{O}}/c \gg h_{\chi}/c$, we can evaluate the correlator by monodromy method, which can be used for the following configuration of operators on the plane \cite{Fitzpatrick:2014vua},
\begin{equation}
\langle\chi(0)\chi(x)\mathcal{O}(1)\mathcal{O}(\infty)\rangle = \lim_{z\rightarrow \infty} z^{2h_{\mathcal{O}} } \langle\chi(0)\chi(x)\mathcal{O}(1)\mathcal{O}(z)\rangle\label{mono-ans}
\end{equation}
Where $x$ is the cross ratio. Via map to the cylinder $z=e^{w}$ this can be related to the light two point function evaluated in the heavy background, which is useful for investigating thermalization and related issues \cite{Fitzpatrick:2015foa, Faulkner:2017hll}. We are however concerned with evaluating the two point function of $\mathcal{O}$ in the light $\chi$ background, hence we are in need a conformal transformation that takes the above to 
\begin{equation}
\langle\chi(0) \chi(\infty) \mathcal{O}(1) \mathcal{O}(1-x)\rangle =  \lim_{z'\rightarrow \infty}z'^{ 2h_{\chi} } \langle\chi(0) \chi(z') \mathcal{O}(1) \mathcal{O}(1-x)\rangle. \label{to}
\end{equation}
It is straight forward to see that the required conformal transformation is given by,
\begin{equation}
\label{eq:conf-transf}
z^{\prime}=\frac{(1-x)z}{z-x} 
\end{equation}
Using eq. \eqref{eq:conf-transf}, the two plane four-point functions of primaries are related as follows, \begin{eqnarray}
\label{trafo}
\langle\chi(0) \chi(\infty) \mathcal{O}(1)\mathcal{O}(1-x)\rangle&=&   \bigg(\frac{\partial z^{\prime}}{\partial z}\bigg)^{-h_{\chi}}_{z^{\prime}=0}\bigg(\frac{\partial z^{\prime}}{\partial z}\bigg)^{-h_{\chi}}_{z^{\prime}\rightarrow \infty }\bigg(\frac{\partial z^{\prime}}{\partial z}\bigg)^{-h_{\mathcal{O}}}_{z^{\prime}=1}\bigg(\frac{\partial  z^{\prime}}{\partial z}\bigg)^{-h_{\mathcal{O}}}_{z^{\prime}=1-x}\nonumber \\
&& \langle\chi(0)\chi(x)\mathcal{O}(1)\mathcal{O}(\infty)
\nonumber \\
&=&x^{2(h_{\chi}-h_{\mathcal{O}})}\lim_{z,z'\rightarrow \infty} z'^{-2h_{\chi}}z^{2h_{\mathcal{O}}} \langle\chi(0)\chi(x)\mathcal{O}(1)\mathcal{O}(\infty)\rangle\nonumber\\
\end{eqnarray} 
and we arrive at the equation
\begin{equation}
\lim_{z^{\prime}\rightarrow\infty}z^{2h_{\chi}}\langle\chi(0) \chi(z^{\prime})\mathcal{O}(1)\mathcal{O}(1-x)\rangle=x^{2(h_{\chi} - h_{\mathcal{O}})}\lim_{z \rightarrow \infty} z^{2h_{\mathcal{O}}}\langle\chi(0)\chi(x)\mathcal{O}(1) \mathcal{O}(z)\rangle.  
\end{equation}
In deriving the above we used  $\frac{\partial z'}{\partial z}=\frac{x(x-1)}{(z-x)^{2}}$. It is also to be noted that $\lim_{z'\rightarrow 1-x}=\lim_{z\rightarrow\infty}$. Assuming identity block dominance \footnote{Using Tauberian theorems for \textit{sufficiently} heavy $\Delta_{\mathcal{O}}$ one can make the identity dominance \cite{Das:2020uax} assumption a bit more precise.}, one thereby has \cite{Fitzpatrick:2014vua},
 \begin{equation} 
 \langle\chi(0)\chi(\infty)\mathcal{O}(1)\mathcal{O}(1-x)\rangle=x^{2(h_{\chi}-h_{\mathcal{O}})}\left(\frac{1 - (1-x)^{\alpha_{\mathcal{O}}} }{\alpha_{\mathcal{O}}} \right)^{-2h_{\chi}} ( 1-x)^{- h_{\chi}(1-\alpha_{\mathcal{O}})},  \label{eq:identity-dom}
\end{equation} 
where, $\alpha_{\mathcal{O}}=\sqrt{1-\frac{12 \Delta_{\mathcal{O}}}{c}}$. Finally we need to relate the expression of eq. \eqref{eq:identity-dom} for plane correlator with that on cylinder via the transformation $w=\log z$. This is given by,\\ 
\begin{equation}
\langle\chi\vert\mathcal{O}(w)\mathcal{O}(0)\vert\chi\rangle =(1-x)^{h_{\mathcal{O}}}\langle\chi(0)\mathcal{O}(1) \mathcal{O}(1-x)\chi(\infty)\rangle. 
\end{equation}\\
Note that in the monodromy computation we ignore $\mathcal{O}(\Delta_{\chi}^{2}/c^{2} )$. 
Next we we write the cylinder correlator in terms of the complex coordinate on the cylinder and get,
\begin{equation}
\langle\chi \vert \mathcal{O}(0) \mathcal{O}(w) \vert \chi\rangle=2^{-2h_{\mathcal{O}}} \alpha_{\mathcal{O}}^{2 h_{\chi}} \frac{ \sinh^{2(h_{\chi} - h_{\mathcal{O}} )}\frac{w}{2} }{\sin^{2h_{\chi}} \frac{w\alpha_{\mathcal{O}}}{2}}. 
\end{equation}
Note that this answer goes over to the leading piece when $\chi$ is the identity operator. 
The thermal two point function, with the correction term, now looks like  
\begin{equation}
\langle\mathcal{O}(t) \mathcal{O}(0)\rangle_{\beta} \propto \left(\frac{\pi}{\beta}\right)^{2\Delta_{\mathcal{O}}} \frac{1}{Z\left(\frac{4\pi^{2}}{\beta}\right)} \frac{1}{\sinh^{2\Delta_{\mathcal{O}}}\frac{\pi t}{\beta}}\bigg\{1+\sum_{N}g_{N}e^{-\frac{4\pi^{2}N}{\beta}} +\sum_{\chi}g_{\chi}\alpha_{\mathcal{O}}^{2\Delta_{\chi}}\frac{\sinh^{2\Delta_{\chi}}\frac{\pi t}{\beta}}{\sinh^{2\Delta_{\chi}}\frac{\pi t\alpha_{\mathcal{O}}}{\beta}} \bigg\}.
\end{equation} 
If we neglect terms beyond $\mathcal{O}(\Delta_{\chi}/c)$ the correction coming from the third term can be incorporated in the inverse temperature $\beta$
\begin{eqnarray}  
\label{compact}
\frac{\alpha_{\mathcal{O}}^{2\Delta_{\chi}}}{\beta^{2\Delta_{\mathcal{O}}}} \frac{\sinh^{2(\Delta_{\chi}-\Delta_{\mathcal{O}})}\frac{\pi t}{\beta}}{\sinh^{2\Delta_{\chi}}\frac{\pi t\alpha_{\mathcal{O}}}{\beta}}&=&\frac{(1-12\Delta_{\chi}\Delta_{\mathcal{O}}/c)}{\beta^{2\Delta_{\mathcal{O}}}\sinh^{2\Delta_{\mathcal{O}}}\frac{\pi t}{\beta}}\frac{\sinh^{2\Delta_{\chi}}\frac{\pi t}{\beta}}{\sinh^{2\Delta_{\chi}}\frac{\pi t}{\beta}(1-6\Delta_{\mathcal{O}}/c)} \nonumber\\
&\approx&\frac{\alpha_{\chi}^{2\Delta_{\mathcal{O}}}}{\beta^{2\Delta_{\mathcal{O}}}}\frac{\sinh^{-2\Delta_{\mathcal{O}}}\frac{\pi t}{\beta}}{\left(1-\frac{6\pi t\Delta_{\mathcal{O}}}{c\beta}\coth\frac{\pi t}{\beta}\right)^{2\Delta_{\chi}}}\nonumber\\
&\approx&\beta_{\chi}^{-2\Delta_{\mathcal{O}}} \frac{1}{\sinh^{2\Delta_{\mathcal{O}}}\frac{\pi t}{\beta_{\chi}}} 
\end{eqnarray} 

Here, 
\begin{eqnarray}
\label{temp_corrc}
\beta_{\chi}&=&\beta\left(1-\frac{12\Delta_{\chi}}{c}\right)^{-1/2}\nonumber\\
&=&\beta\left(1+\frac{6\Delta_{\chi}}{c}\right)
\end{eqnarray}

\subsection{$c\to\infty$, $h_{\mathcal{O}}$ and $h_{\chi}$ held fixed} 
The four point function in CFT can be written as sum over a complete set of exchanged states.
\begin{equation}
 \langle\mathcal{O}(\infty)\mathcal{O}(1)\mathcal{O}(z,\bar{z})\mathcal{O}(0)\rangle\approx\sum\langle\mathcal{O}(\infty)\mathcal{O}(1)\frac{\vert\alpha\rangle\langle \alpha\vert}{\langle\alpha\vert\alpha\rangle}\mathcal{O}(z,\bar{z})\mathcal{O}(0)\rangle    
\end{equation}
With $c\to\infty$ and the dimensions of the external operators held fixed the construction of the complete set is simpler. Upto leading order in $\left(1/c\right)$ one only needs to consider the descendants obtained by acting with $L_{-1}$. The norm of the states $L_{-n}\vert h\rangle$ for $n\geq2$ are proportional to $c$ hence has only a sub-leading contribution to the virasoro blocks. The projector for the leading block contribution can be approximated by \cite{Fitzpatrick:2015zha}
\begin{equation}
    \mathcal{P}_{h,\bar{h}}\approx\sum_{k}\frac{ L_{-1}^{k}\vert h\rangle \langle h\vert L_{1}^{k}}{\langle h\vert L_{1}^{k}L_{-1}^{k}\vert h\rangle},
\end{equation}
To get the $\mathcal{O}\left(\frac{1}{c}\right)$ correction one needs to construct $\mathcal{P}_{h,\bar{h}}$ accordingly. For light external operators ($\Delta_{i}$) this has been carried out with considerable mathematical rigour in \cite{Bombini:2018jrg} for $\langle\chi(\infty)\chi(1)\mathcal{O}(x)\mathcal{O}(0)\rangle$ configuration. This can be connected to $\langle\chi(\infty)\chi(0)\mathcal{O}(1)\mathcal{O}(1-x)\rangle$ via the transformation $z\rightarrow\frac{x-1}{z-1}$. The expression for vacuum block was derived in \cite{Perlmutter:2015iya} and can also be independently verified from the result for general $h$ in \cite{Bombini:2018jrg}   
\begin{eqnarray}
\langle\chi\vert\mathcal{O}(t_{euc}/\tau)\mathcal{O}(0)\vert\chi\rangle&=&(1-x)^{h_{\mathcal{O}}}\langle\chi(0)\mathcal{O}(1)\mathcal{O}(1-x)\chi(\infty)\rangle \nonumber\\
&=&(1-x)^{h_{\mathcal{O}}}x^{-2h_{\mathcal{O}}}\left[ 1+\frac{2h_{\chi}h_{\mathcal{O}}x^{2}}{c} {_{2}}F_{1}(2,2,4,x)\right]\nonumber\\
&=&\left(\sinh\frac{\pi t}{\beta}\right)^{-2h_{\mathcal{O}}}\left[1-\frac{24 h_{\chi}h_{\mathcal{O}}}{c}\left(1-\frac{\pi t}{\beta}\coth\frac{\pi t}{\beta}\right)\right]\nonumber\\ \nonumber
\end{eqnarray}
Together with the anti-holomorphic part, up to $\mathcal{O}\left(\frac{\Delta_{\chi}}{c}\right)$ the correction term takes the following form
\begin{eqnarray}
    &&\frac{\alpha_{\chi}^{2\Delta_{\mathcal{O}}}}{\beta^{2\Delta_{\mathcal{O}}}}\frac{1}{\sinh^{2\Delta_{\mathcal{O}}}\left(\frac{\pi t}{\beta}(1-12\Delta_{\chi}/c)^{1/2}\right)}\nonumber\\
    &=&\frac{1}{\beta_{\chi}^{2\Delta_{\mathcal{O}}}}\frac{1}{\sinh^{2\Delta_{\mathcal{O}}}\frac{\pi t}{\beta_{\chi}}}
\end{eqnarray}\\
\vspace{2mm}\\
For the two regimes of the dimensions $\Delta_{\mathcal{O}}$ and $\Delta_{\chi}$ that we are primarily interested in, till order $\Delta_{\chi}/c$, the finite temperature correction term can be written as $\mathcal{O}$ correlator at a rescaled temperature $\beta_{\chi}$. This is our main result. We shall show in next section how this change in temperature drives the change in frequency of quasinormal modes. It may be of interest to mention here that similar expression of thermal two point function at rescaled temperature has been worked out in \cite{David:2019bmi} for lighter operators inserted inside heavy states. On the bulk side this can be interpreted as the correlation function of $\mathcal{O}$ in a spacetime that gets perturbative correction from the $\chi$ fields.\\ 
It turns out that in the perturbative regime of $\Delta_{\chi}/c$ the dimension of the probe operator does not influence the change in temperature. We have the same $\beta_{\chi}$ as in eq. \eqref{temp_corrc}. The change in temperature is same in both the cases when $\Delta_{\mathcal{O}}\gg\Delta_{\chi}$ and when there is no dimensional hierarchy between the operators. 

\section{Correction to QNMs}
\label{corrc}
As discussed in section \ref{sec:intro} the rate of decay/emission of the black hole can be obtained by performing Fourier transform of the autocorrelation of $\mathcal{O}$ (operator dual to the bulk probe scalar) in the thermal state of the CFT. Due to the inclusion of the contribution of the light states, now we have a series of QNMs for each light operator. In the regime we are working in the effect of the light states can be incorporated as shift in temperature. We find the new set of QNMs by tracking the change in the location of the poles of gamma function in eq. \eqref{decay2}. 
\begin{eqnarray}
\label{qnmC}\omega_{n}^{\chi} = - \frac{2\pi i } {\beta_{\chi}} (n+\Delta_{\mathcal{O}} ) = - \frac{2\pi i }{\beta} \left( n + \Delta_{\mathcal{O}}\right)\left(1-\frac{6\Delta_{\chi}}{c} \right) 
\end{eqnarray}
The leading answer, eq. \eqref{qnm_mode} can be viewed in this light, as the QNMs associated with just pure gravity. \\ 
  Stability demands that the poles be situated in the lower half plane. This implies that $\Delta_{\chi} < c/6$. This is consistent for our purpose since the motivation for studying the finite temperature corrections on the boundary lies in studying the effect of light matter fields on the bulk. This also satisfies Hellerman's bound on the lightest non-vacuum primary \cite{Hellerman:2009bu}. If $\Delta_{\chi}$ touches $c/6$ then we enter \textit{non-perturbative} regime, and we cannot use anymore the HLLH monodromy answer.\\

\section{Discussions}
\label{discussion}
The correction term arising due to finite temperature expansion of torus two point function gets attributed to the presence of light matter field along with gravity in the bulk. Up to effects of the  $\mathcal{O}(\Delta_{\chi}/c)$ the correction is encoded in the change of inverse temperature $\beta$ to $\beta\left(1+\frac{6\Delta_{\chi}}{c}\right)$. In other words, the probe particle `feels' that only gravity is turned on but the system temperature is different than that of the blackhole temperature. For a unitary theory ($\Delta_{\chi}, c > 0$) this temperature $T_{\chi}$ is less than blackhole temperature $T$ with no matter field present. Another way to interpret it is by looking at the change in outer radius of BTZ blackhole. BTZ temperature is related to the dimension of the CFT state in the boundary by $2\pi T = (12\Delta/c-1)^{1/2}$. Since radius of the outer horizon of BTZ is proportional to $\sqrt{\Delta}$, change in temperature affects the blackhole geometry.          
Next, the expression for the thermal two point function of the probe operator $\mathcal{O}$ can be interpreted as the geodesic length of the probe particle in the BTZ background \cite{Louko:2000tp}. It will be interesting to look for explanations of the change in temperature by studying the geodesic length in presence of backreacting matter fields in order to recover the observed holographic corrections. \\ 

As discussed in \cite{Kraus:2016nwo} a semi-classical AdS interpretation is possible even if one does not take $c\to\infty$ in the boundary theory. It might be interesting to study the scenario when the mass of the matter field is well outside the perturbative regime $(\Delta_{\chi}\gtrsim c/6)$ but still not infinite. In this case $\vert\chi\rangle$ is no longer dual to a perturbative field in the bulk, rather it backreacts on the AdS geometry to create a conical deficit.\\       
In the appendix we discuss the nature of the correction term if we do not restrict ourselves only to the vacuum contribution of the conformal blocks. The expressions are non-trivial and the effect of matter fields does not manifest as change of temperature. An interesting way forward might be trying to explain them along the line of arguments as discussed in \cite{Hijano:2015rla}. 
\section{Acknowledgements}
The author would like to express her gratitude to Dr. Diptarka Das and Dr. Pinaki Banerjee for helping in crystallizing the problem statement and useful comments on the manuscript. The discussions with Dr. Suchetan Das have proven to be illuminating. Dr. Nilay Kundu's comments on the draft have given the author the clarity needed to structure the paper in the current form.     
\appendix
\section{Leading term in high temperature expansion} \label{ap:leading-high-temp}
The conformal transformation we use to take us from the spatial cylinder ($w$) to the plane ($z$) is the exponential map, $z=e^{w}$. We know the two point function on plane is given by, 
\begin{equation}
\langle\mathcal{O}_{1}(z_{1},\bar{z}_{1})\mathcal{O}_{2}(z_{2},\bar{z}_{2})\rangle=\frac{\delta_{12}}{|z_{1}-z_{2}|^{2h_{1}}|\bar{z}_{1}-\bar{z}_{2}|^{2\bar{h}_{1}}}
\end{equation}
The $n$-point functions on plane and cylinder are related by the following transformation.
\begin{eqnarray}
\langle\mathcal{O}_{1}(w_{1},\bar{w}_{1})....\mathcal{O}_{n}(w_{n},\bar{w}_{n})\rangle
&=&\prod\left(\frac{\partial w}{\partial z}\right)^{-h_{i}}_{w=w_{i}}\left(\frac{\partial\bar{w}}{\partial \bar{z}}\right) ^{-\bar{h_{i}}}_{\bar{w}=\bar{w}_{2}}\nonumber\\
&&\langle\mathcal{O}_{1}(z_{1},\bar{z}_{1})....\mathcal{O}_{n}(z_{n},\bar{z}_{n})\rangle
\end{eqnarray}
$h_{i}$ is conformal dimension of $\mathcal{O}_{i}$. Using the above, we find: 

\begin{eqnarray}
w&=&\log z\\ \nonumber
\frac{\partial w}{\partial z}&=&\frac{1}{z}\\ \nonumber 
&=& e^{-w} \\ \nonumber
\langle\mathcal{O}(w_{1},\bar{w}_{1})\mathcal{O}(w_{2},\bar{w}_{2})\rangle&=&\frac{e^{h(w_{1}+w_{2})}e^{\bar{h}(\bar{w}_{1}+\bar{w}_{2})}}{(e^{w_{1}}-e^{w_{2}})^{2h}(e^{\bar{w}_{1}}-e^{\bar{w}_{2}})^{2\bar{h}}}\\
&=&\frac{2^{-2(h+\bar{h})}}{\left( \sinh\frac{(w_{1}-w_{2})}{2}\right) ^{2h}\left(\sinh\frac{\bar{w}_1-\bar{w}_{2}}{2}\right)^{2\bar{h}}} \label{plcy} \nonumber\\
\end{eqnarray}
Where $h$ is the conformal dimension of the operator $\mathcal{O}$. The leading term from eq. \eqref{towork} is, 
\begin{equation}
\langle\mathcal{O}(0,0)\mathcal{O}(w_{2},\bar{w}_{2})\rangle \approx \left(\frac{2\pi}{\beta}\right)^{2\Delta_{\mathcal{O}}}(i)^{-2s_{\mathcal{O}}} \langle\mathcal{O}(0,0) \mathcal{O}(w_{2}^{\prime},\bar{w}_{2}^{\prime})\rangle  \label{set}
\end{equation}

Using eq. \eqref{plcy} we have, 
\begin{eqnarray}
 \langle\mathcal{O}(0,0) \mathcal{O}(w_{2}^{\prime} , \bar{w}_{2}^{\prime})&=& \frac{2^{-2\Delta_{\mathcal{O}} }e^{\frac{\pi^{2}c}{3\beta}}}{\sinh^{2h_{\mathcal{O}}} (- \frac{w_{2}^{\prime}}{2})\sinh^{2\bar{h}_{\mathcal{O}}}(- \frac{\bar{w}_{2}^{\prime}}{2})}, \nonumber \\
 &=&\frac{(-1)^{2\Delta_{\mathcal{O}}}2^{-2\Delta_{\mathcal{O}} }e^{\frac{\pi^{2}c}{3\beta}}}{\sinh^{2h_{\mathcal{O}}}  \frac{w_{2}^{\prime}}{2}\sinh^{2\bar{h}_{\mathcal{O}}} \frac{\bar{w}_{2}^{\prime}}{2}}.
\end{eqnarray}
Next, the modular transformation eq. \eqref{set} gives, 
\begin{equation}
\langle\mathcal{O}(0)\mathcal{O}(t_{euc})\rangle \approx (-1)^{\Delta_{\mathcal{O}}}\left(\frac{\pi }{\beta} \right)^{2\Delta_{\mathcal{O}}} \frac{ e^{\frac{\pi^2 c}{3 \beta } } }{ \sinh^{2h_{\mathcal{O}} } \frac{\pi t_{euc}}{i \beta} \sinh^{2\bar{h}_{\mathcal{O}} } \frac{ \pi t_{euc} }{-i \beta } }. 
\end{equation}   
 Next, we analytically continue, $t_{euc} = it$ and normalize by the leading high temperature partition function, $Z(\beta) = e^{\pi^2 c /(3\beta) }$ to get, 
 $ (-1)^{\Delta_{\mathcal{O}}} \left(\frac{\pi }{\beta} \right)^{2\Delta_{\mathcal{O}}}\frac{ 1}{ \sinh^{2\Delta_{\mathcal{O}}}\frac{\pi t}{ \beta}}, 
 $
 which is the leading order integrand in eq. \eqref{limit}. 
 
 \section{Other block contributions} 
 \label{other-block}
 So far while computing corrections to thermal auto-correlator we assumed identity dominance (see eq. \eqref{eq:identity-dom}). Here for the sake of completeness we take into account the contribution from non-identity block. 
 \subsection{Non-identity block}
 Our first case for consideration is the expression for the virasoro block derived by monodromy method in \cite{Fitzpatrick:2014vua}. We further assume that the dimension of the exchanged operator, $\Delta_{p}$ does not scale with $c$ and $\Delta_{p}\ll 1$.   

\begin{equation}
\left\langle\mathcal{O}(1)\mathcal{O}(\infty)\chi(0)\chi(x)\right\rangle=(1-x)^{-h_{\chi}(1-\alpha_{\mathcal{O}})}\left(\frac{1-(1-x)^{\alpha_{\mathcal{O}}}}{\alpha_{\mathcal{O}}}\right)^{(h_{p}-2h_{\chi})} \left(\frac{1+(1-x)^{\alpha_{\mathcal{O}}/2}}{2}\right)^{-2h_{p}}
\end{equation} 
Using the map eq. \eqref{trafo} we get the desired configuration of operators on cylinders. 
\begin{eqnarray}
 \left\langle\chi|\mathcal{O}(0)\mathcal{O}(\omega)|\chi\right\rangle&=& x^{2(h_{\chi}-h_{\mathcal{O}})}(1-x)^{h_{\mathcal{O}}}\left\langle\mathcal{O}(0)\mathcal{O}(x)\chi(1)\chi(\infty)\right\rangle\nonumber\nonumber\\
 &=&4^{(h_{p}-h_{\mathcal{O}})}\alpha_{\mathcal{O}}^{(2h_{\chi}-h_{p})}\frac{\left(\sinh\frac{\omega}{2}\right)^{2(h_{\chi}-h_{\mathcal{O}})}}{\left(\sinh\frac{\omega\alpha_{O}}{2}\right)^{2h_{\chi}}}\left(\tanh\frac{\omega\alpha_{\mathcal{O}}}{4}\right)^{h_{p}}\nonumber\\
 \end{eqnarray}
Plugging in the anti-holomorphic part and analytically to realtime domain via $w \to \frac{t_{euc}}{\tau}$,
 \begin{eqnarray}
 \left\langle\chi|\mathcal{O}(0)\mathcal{O}(t_{euc}/\tau,t_{euc}/\bar{\tau} )|\chi\right\rangle_{non-id} &\approx& \alpha_{\mathcal{O}}^{(2\Delta_{\chi}-\Delta_{p})}\frac{\left(\sinh\frac{\pi t}{\beta}\right)^{2(\Delta_{\chi}-\Delta_{\mathcal{O}})}}{\left(\sin\frac{\alpha_{O}\pi t}{\beta}\right)^{2\Delta_{\chi}}}\left(\tanh\frac{\alpha_{\mathcal{O}}\pi t}{2\beta}\right)^{\Delta_{p}} \nonumber\\
 &=&\left(\frac{4}{\alpha_\mathcal{O}} \tanh\frac{\alpha_{\mathcal{O}}\pi t}{2\beta}\right)^{\Delta_{p}} \times   \left\langle\chi|\mathcal{O}(0)\mathcal{O}(t_{euc}/\tau,t_{euc}/\tau )|\chi\right\rangle_{id}\nonumber\\
 \end{eqnarray}
 The effect of $\mathcal{O}_p$ is incorporated in as a multiplicative factor 
 $f_p = \left(\frac{4}{\alpha_\mathcal{O}} \tanh\frac{\alpha_{\mathcal{O}}\pi t}{2\beta}\right)^{\Delta_{p}}$ to the cylinder correlator $\langle \chi| \mathcal{O}(0) \mathcal{O}(t) |\chi \rangle$.
 To keep only the leading correction coming from the non-identity exchange we keep only terms linear in $\Delta_p$. In that approximation $f_p$ reduces to
 	\begin{equation}
 	f_{p} = 1 +  \Delta_{p} \left[\log \left(4 \tanh \left(\frac{\pi  t}{2 \beta }\right)\right) +  \frac{6 \Delta_{\mathcal{O}}}{c}  \left(1-\frac{\frac{\pi  t}{\beta}}{  \sinh \left(\frac{\pi  t}{\beta }\right)}\right) \right].
 	\end{equation} 
The contribution from the exchanged operator $\Delta_{p}$ has a complicated form and can't be expressed simply in terms of change in temperature as in the case of vacuum block.

 \subsection{Heavy Exchange} \label{ap:heavy-exchange}
In this section we consider another regime, where the expression for the virasoro block is known, namely when the exchanged operator is very heavy ie $\Delta \gg \Delta_{\mathcal{O}}$, $\Delta_{\chi}$ and it takes the form \cite{Besken:2019jyw}, 
 \begin{equation}
 \left\langle\chi(0)\chi(x)\mathcal{O}(1)\mathcal{O}(\infty)\right\rangle=x^{-h_{\chi}-h_{\mathcal{O}}}(16q)^{h},
 \end{equation}
 where, $q=\exp\left[-\pi\frac{K(1-x)}{K(x)}\right]$ and $x$ is the modulus with $0 \le x < 1.$ Note that this is very different regime in contrast to the ones we have considered above so far. Therefore we shouldn't expect to obtain similar modifications to the thermal 2-point function. We can expand $q$ near small $x$,
 \begin{equation}
 q=\frac{x}{16}+\frac{x^{2}}{32}+\frac{21x^{3}}{1024} + \ldots
 \end{equation}
 Keeping only upto the first order in $x$ and using the conformal transformation in eq. \eqref{eq:conf-transf}, we get
 \begin{equation}
 \left\langle\chi|\mathcal{O}(0)\mathcal{O}(t_{euc}/\tau, t_{euc}/\bar{\tau})|\chi\right\rangle=(-2)^{\Delta+\Delta_{\chi}-3\Delta_{\mathcal{O}}}\left(\sinh\frac{\pi t}{\beta}\right)^{\Delta+\Delta_{\chi}-3\Delta_{\mathcal{O}}}
 \end{equation}
If we scale $\Delta_\chi$ and $\Delta_{\mathcal{O}}$ with $\epsilon$ and expand in small $\epsilon$, the RHS yields,
 	\begin{align}
 	\left(-2 \sinh \frac{\pi  t}{\beta }\right)^{\Delta } \left[1+ (\Delta_\chi -3 \Delta_{\mathcal{O}}) \log \left(-2 \sinh \left(\frac{\pi  t}{\beta }\right)\right)\right]
 	\end{align}
 The leading behaviour of the correlator is $\left(-2 \sinh \frac{\pi  t}{\beta }\right)^{\Delta }$ which is quite different from the integrand in eq. \eqref{limit}. This is not unexpected since the exchanged operator is dominating and at leading order we drop both  $\Delta_{\chi}$ and $\Delta_{\mathcal{O}}$.

\section{Contribution from the descendants of vacuum} \label{ap:vacuum desc} 
At level $N$ the contribution from the descendant states of vacuum to the high temperature expansion of the torus two point function takes the following form,
\begin{align}
\frac{\langle 0\vert L_{m_{n}}^{k_{n}}..L_{m_{1}}^{k_{1}}\mathcal{O}(w_{1})\mathcal{O}(w_{2})L_{-m_{1}}^{k_{1}}..L_{-m_{n}}^{k_{n}}\vert 0\rangle}{\langle 0\vert L_{m_{n}}^{k_{n}}..L_{m_{1}}^{k_{1}}L_{-m_{1}}^{k_{1}}..L_{-m_{n}}^{k_{n}}\vert 0\rangle} 
\end{align} 
Where $\sum_{i}m_{i}k_{i}=N$. At level $N=2$, 
\begin{eqnarray}
\label{descvac}
\frac{\langle 0\vert L_{2}\mathcal{O}(w_{1})\mathcal{O}(w_{2})L_{-2}\vert 0\rangle}{\langle 0\vert [L_{2},L_{-2}]\vert 0\rangle}&=&\frac{\langle 0\vert[L_{2},\mathcal{O}(w_{1})]\mathcal{O}(w_{2})L_{-2}\vert 0\rangle}{\langle 0\vert [L_{2},L_{-2}]\vert 0\rangle} + \frac{\langle 0\vert\mathcal{O}(w_{1})[L_{2},\mathcal{O}(w_{2})]L_{-2}\vert 0\rangle}{\langle 0\vert [L_{2},L_{-2}]\vert 0\rangle} \nonumber\\
&+&\frac{\langle 0\vert\mathcal{O}(w_{1})\mathcal{O}(w_{2})[L_{2},L_{-2}]\vert 0\rangle}{\langle 0\vert [L_{2},L_{-2}]\vert 0\rangle}
\end{eqnarray}
The commutators between $\mathcal{O}$ and the virasoro generator $L_{n}$ are evaluated via 
\begin{align}
\label{commu}
[L_{n},\mathcal{O}(w)]= h(n+1)w^{n}\mathcal{O}(w)+w^{n+1}\partial\mathcal{O}(w)     
\end{align}
The first and second term thus together becomes
\begin{eqnarray}
\label{laurnt}
(3h w_{1}^{2}+w_{1}^{3}\partial_{w_{1}}+3h w_{2}^{2}+w_{2}^{3}\partial_{w_{2}})\langle 0\vert\mathcal{O}(w_{1})\mathcal{O}(w_{2})L_{-2}\vert 0\rangle  
\end{eqnarray}
To evaluate the correlation function in \ref{laurnt} we use the integral expression of $L_{n}$ in terms of stress energy tensor $T(z)$.
\begin{equation}
    \langle 0\vert\mathcal{O}(w_{1})\mathcal{O}(w_{2})L_{-2}\vert 0\rangle=\frac{1}{2\pi i}\oint z^{-1}\langle \mathcal{O}(w_{1})\mathcal{O}(w_{2})T(z)\rangle dz 
\end{equation} 
The correlation function of the primaries with the stress tensor can be expressed as the correlation function of the primaries (\cite{DiFrancesco:1997nk}, \cite{Polchinski:1998rr}) via  
\begin{align}
    \langle T(w)\mathcal{O}_{1}(w_{1}) \mathcal{O}_{2}(w_{2})...\mathcal{O}_{n}(w_{n})\rangle =\left(\sum_{i}^{n}\frac{h_{i}}{(w-w_{i})^{2}} +\frac{1}{(w-w_{i})}\frac{\partial}{\partial w_{i}}\right)\langle\mathcal{O}_{1}(w_{1})\mathcal{O}_{2}(w_{2})...\mathcal{O}_{n}(w_{n})\rangle   
\end{align} 
However a more generalized treatment would be considering the term $\langle\mathcal{O}(w_{1})\mathcal{O}(w_{2})L_{-n}\rangle$ for $n>0$. Following the same steps as above we are required to evaluate two contour integrations. 
\begin{align}
\label{contour}
\oint \frac{w^{1-n}}{(w-w_{i})^{2}}dw\hspace{2mm} \oint \frac{w^{1-n}}{(w-w_{i})}dz
\end{align}
From here it is easily verifiable that the above contour integrations give null results. The only non-zero coontribution in \eqref{descvac} comes from the last term, $\langle 0\vert\mathcal{O}(w_{1})\mathcal{O}(w_{2})[L_{2},L_{-2}]\vert 0\rangle$. 
\begin{align}
    \langle\mathcal{O}(w_{1})\mathcal{O}(w_{2})[L_{2},L_{-2}]\rangle=\frac{c}{2}\langle\mathcal{O}(w_{1})\mathcal{O}(w_{2})\rangle
\end{align}
The multiplicative factor of $\frac{c}{2}$ gets cancelled by the normalization factor in the denominator. 
The calculation gets more involved if one deals with a descendant state of the form $L_{-m_{1}}^{k_{1}}L_{-m_{2}}^{k_{2}}...L_{-m_{n}}^{k_{n}}\vert 0\rangle$. The prescription is to take all the $L_{n}$' s on the left to the right side of the operators $\mathcal{O}(w_{1})$ and $\mathcal{O}(w_{2})$ by applying successive commutations between a generator and an operator. The only nonzero contribution comes from the term
\begin{align}
   \frac{ \langle 0\vert\mathcal{O}(w_{1})\mathcal{O}(w_{2})L_{m_{n}}^{k_{n}}..L_{m_{1}}^{k_{1}}L_{-m_{1}}^{k_{1}}..L_{-m_{n}}^{k_{n}}\vert 0\rangle}{\langle 0\vert L_{m_{n}}^{k_{n}}..L_{m_{1}}^{k_{1}}L_{-m_{1}}^{k_{1}}..L_{-m_{n}}^{k_{n}}\vert 0\rangle} 
\end{align}
It is evident from the above expression that one is required to calculate the same commutators between $L_{n}$'s in the numerator and the denominator, cancelling out the expression containing the central charge $c$\hspace{2mm}. What is remaining to be shown is the fate of the terms like $\langle 0\vert L_{m_{n}}^{k_{n}}..[L_{m_{p}}^{k_{p}},\mathcal{O}(w_{1})]\mathcal{O}(w_{2})L_{m_{p-1}}^{k_{p-1}}..L_{m_{1}}^{k_{1}}L_{-m_{1}}^{k_{1}}..L_{-m_{n}}^{k_{n}}\vert 0\rangle$. For terms like this, the commutator is replaced with eqn. \eqref{commu} and the $L_{n}$'s are taken to the right successively. We are now left with integrations of the form of eq. \eqref{contour}. Though a more convincing and robust proof of the preceding argument is not available it has hold true for the levels checked (up to $N=5$) so far. While we have discussed only about the diagonal terms the argument extends to non diagonal terms like $\langle 0\vert L_{m_{n}}^{k_{n}}..L_{m_{1}}^{k_{1}}\mathcal{O}(w_{1})\mathcal{O}(w_{2})L_{-n_{1}}L_{-n_{2}}\vert 0\rangle$, where $n_{1}+n_{2}=\sum_{i}^{n}m_{i}k_{i}=N$.  
\bibliographystyle{unsrt}

\end{document}